\documentclass[conference,final,twoside]{IEEEtran}

\usepackage[hyphens]{url}
\usepackage{lscape}
\usepackage{amsmath}

\usepackage{xcolor}
\definecolor{codecolor}{rgb}{0.94, 1.0, 1.0}
\definecolor{commentcolor}{rgb}{0.54, 0.17, 0.89}
\definecolor{keywordcolor}{rgb}{0.28, 0.02, 0.03}
\definecolor{stringcolor}{rgb}{0.0, 0.13, 0.28}

\usepackage{listings}

\usepackage{relsize}

\newtheorem{example}{Example}

\newcommand*{\fund}[3]{\mathord{#1}\colon#2\to#3}

\newcommand*{\q}{$\mathord{\circ}$}

\providecommand*{\Pset}{\mathbb{P}}            
\providecommand*{\Cset}{\mathbb{C}}            
\providecommand*{\Iset}{\mathbb{I}}            
\providecommand*{\Nset}{\mathbb{N}}            

\newenvironment{nbquote}
 {\quote\interlinepenalty=10000 }
 {\endquote}

\newcommand{\codesize}{\smaller[1.5]}
\lstdefinestyle{Cfamily}{%
basicstyle=\ttfamily\codesize,
commentstyle=\color{commentcolor},
moredelim=[is][{\btHL[onslide=<2->{fill=red!30}]}]{/*!*/}{/*!*/},
moredelim=[is][{\btHL[onslide=<1->{fill=green!30}]}]{/*=*/}{/*=*/},
keywordstyle=\color{keywordcolor},
identifierstyle=,
stringstyle=\color{stringcolor},
showstringspaces=false
}

\lstdefinestyle{inlineCfamily}{%
basicstyle=\ttfamily,
commentstyle=\color{commentcolor},
keywordstyle=\color{keywordcolor},
identifierstyle=,
stringstyle=\color{stringcolor},
showstringspaces=false,
}

\newcommand{\code}[1]{\lstinline[language=C,style=inlineCfamily]@#1@}

\lstnewenvironment{Ccode}{\lstset{
language=C,
style=Cfamily,
frame=single,
aboveskip=10pt,
belowskip=10pt
}}{}

\newcommand{\includeCcode}[1]{\lstinputlisting[
language=C,
style=Cfamily,
frame=single,
aboveskip=10pt,
belowskip=10pt
]{#1}}

\usepackage{afterpage}
\usepackage{enumitem}
\usepackage{flushend}
\usepackage{amssymb}
\usepackage{graphicx}
\usepackage{fancyvrb}
\usepackage{times}
\usepackage[hidelinks]{hyperref}
\usepackage{xcolor}
\hypersetup{
    breaklinks=true,
    colorlinks,
    linkcolor={red!50!black},
    citecolor={blue!50!black},
    urlcolor={blue!80!black}
}

\usepackage{xstring}

\newcommand{\MCId}[1]{%
\IfStrEqCase{#1}{%
{D4.1}{Dir~4.1}
{R1.2}{\textit{Rule~1.2}}
{R1.3}{Rule~1.3}
{R2.1}{Rule~2.1}
{R2.2}{Rule~2.2}
{R8.13}{\textit{Rule~8.13}}
{R9.1}{\textbf{Rule~9.1}}
{R12.2}{Rule~12.2}
{R13.1}{Rule~13.1}
{R13.2}{Rule~13.2}
{R13.5}{Rule~13.5}
{R14.1}{Rule~14.1}
{R14.2}{Rule~14.2}
{R14.3}{Rule~14.3}
{R17.2}{Rule~17.2}
{R17.5}{\textit{Rule~17.5}}
{R17.8}{\textit{Rule~17.8}}
{R18.1}{Rule~18.1}
{R18.2}{Rule~18.2}
{R18.3}{Rule~18.3}
{R18.6}{Rule~18.6}
{R19.1}{\textbf{Rule~19.1}}
{R21.13}{\textbf{Rule~21.13}}
{R21.14}{Rule~21.14}
{R21.17}{\textbf{Rule~21.17}}
{R21.18}{\textbf{Rule~21.18}}
{R21.19}{\textbf{Rule~21.19}}
{R21.20}{\textbf{Rule~21.20}}
{R22.1}{Rule~22.1}
{R22.2}{\textbf{Rule~22.2}}
{R22.3}{Rule~22.3}
{R22.4}{\textbf{Rule~22.4}}
{R22.5}{\textbf{Rule~22.5}}
{R22.6}{\textbf{Rule~22.6}}
{R22.7}{Rule~22.7}
{R22.8}{Rule~22.8}
{R22.9}{Rule~22.9}
{R22.10}{Rule~22.10}}%
[unknown MC3R1 Id #1]%
}

\newcommand*{\MCHd}[1]{%
\IfStrEqCase{#1}{%
{D4.1}{Run-time failures shall be minimized}
{R1.2}{Language extensions should not be used}
{R1.3}{There shall be no occurrence of undefined or critical unspecified behaviour}
{R2.2}{There shall be no \textit{dead code}}
{R8.13}{A pointer should point to a \code{const}-qualified type whenever possible}
{R8.14}{The \code{restrict} type qualifier shall not be used}
{R9.1}{The value of an object with automatic storage duration shall not be read before it has been set}
{R14.1}{A \textit{loop counter} shall not have \textit{essentially floating} type}
{R14.3}{Controlling expressions shall not be invariant}
{R17.2}{Functions shall not call themselves, either directly or indirectly}
{R21.13}{Any value passed to a function in \code{<ctype.h>} shall be representable as an \code{unsigned char} or be the value EOF}
{R21.14}{The Standard Library function \code{memcmp} shall not be used to compare null terminated strings}
{R21.19}{The pointers returned by the Standard Library functions \code{localeconv}, \code{getenv}, \code{setlocale} or, \code{strerror} shall only be used as if they have pointer to const-qualified type}
{R22.1}{All resources obtained dynamically by means of Standard Library functions shall be explicitly released}
{R22.3}{The same file shall not be open for read and write access at the same time on different streams}
{R22.5}{A pointer to a \code{FILE} object shall not be dereferenced}
{R22.7}{The macro \code{EOF} shall only be compared with the unmodified return value from any Standard Library function capable of returning \code{EOF}}
{R22.8}{The value of \code{errno} shall be set to zero prior to a call to an \textit{errno-setting-function}}}%
[unknown MC3R1 Hd #1]%
}

\renewcommand*\descriptionlabel[1]{\hspace\labelsep
                                   \normalfont\bfseries #1}

\makeatletter
\newenvironment{btHighlight}[1][]
{\begingroup\tikzset{bt@Highlight@par/.style={#1}}\begin{lrbox}{\@tempboxa}}
{\end{lrbox}\bt@HL@box[bt@Highlight@par]{\@tempboxa}\endgroup}

\newcommand\btHL[1][]{%
  \begin{btHighlight}[#1]\bgroup\aftergroup\bt@HL@endenv%
}
\def\bt@HL@endenv{%
  \end{btHighlight}%
  \egroup
}
\newcommand{\bt@HL@box}[2][]{%
  \tikz[#1]{%
    \pgfpathrectangle{\pgfpoint{1pt}{0pt}}{\pgfpoint{\wd #2}{\ht #2}}%
    \pgfusepath{use as bounding box}%
    \node[anchor=base west, outer sep=0pt,inner xsep=1pt, inner ysep=0pt, rounded corners=2pt, minimum height=\ht\strutbox+1pt,#1]{\raisebox{1pt}{\strut}\strut\usebox{#2}};
  }%
}
\makeatother

\VerbatimFootnotes
\IEEEoverridecommandlockouts
\begin{document}

\title{Coding Guidelines and Undecidability}
\author{\IEEEauthorblockN{Roberto Bagnara\textsuperscript{*}\thanks{%
{}\textsuperscript{*} Roberto Bagnara
is a member of the \emph{MISRA C Working Group} and of
ISO/IEC JTC1/SC22/WG14, a.k.a.\ the \emph{C Standardization Working Group}.
Patricia M.\ Hill is a member of the \emph{MISRA C++ Working Group}.
Nonetheless, the views expressed in this paper are the authors' and should
not be taken to represent the views of the mentioned working groups
and organizations.}}
\IEEEauthorblockA{University of Parma\\
Parma, Italy\\
Email: roberto.bagnara@unipr.it}
\and
\IEEEauthorblockN{Abramo Bagnara}
\IEEEauthorblockA{BUGSENG\\
Parma, Italy\\
Email: abramo.bagnara@bugseng.com}
\and
\IEEEauthorblockN{Patricia M.\ Hill\textsuperscript{*}}
\IEEEauthorblockA{BUGSENG\\
Parma, Italy\\
Email: patricia.hill@bugseng.com}}

\maketitle

\begin{abstract}
  The C and C++ programming languages are widely used for the
  implementation of software in critical systems.  They are complex
  languages with subtle features and peculiarities that might baffle
  even the more expert programmers.  Hence, the general prescription of
  \emph{language subsetting}, which occurs in most functional safety
  standards and amounts to only using a ``safer'' subset of the language,
  is particularly applicable to them.  Coding guidelines
  are the preferred way of expressing language subsets.
  Some guidelines are formulated in terms of the programming language
  and its implementation only: in this case they are amenable to
  automatic checking.  However, due to fundamental limitations of
  computing, some guidelines are \emph{undecidable}, that is, they
  are based on program properties that no current and future algorithm
  can capture in all cases.  The most mature and widespread coding standards,
  the MISRA ones, explicitly tag guidelines with \emph{undecidable}
  or \emph{decidable}. It turns out that this information is not of
  secondary nature and must be taken into account for a full understanding
  of what the guideline is asking for.  As a matter of fact, undecidability
  is a common source of confusion affecting many users of coding standards
  and of the associated checking tools.
  In this paper, we recall the notions
  of decidability and undecidability in terms that are understandable to
  any C/C++ programmer.
  The paper includes a systematic study of all the undecidable
  MISRA~C:2012 guidelines, discussing the reasons for the
  undecidability and its consequences.
  We pay particular attention to undecidable guidelines that have
  decidable approximations whose enforcement would not overly
  constrain the source code.
  We also discuss some coding guidelines for which compliance is hard,
  if not impossible, to prove, even beyond the issue of decidability.
  Findings and lessons learned are reported along with some concrete
  suggestions to improve the state of the art.
\end{abstract}

\section{Introduction}
\label{sec:introduction}

\emph{Coding guidelines} are restrictions in the way high-level
programming languages can be used to construct programs.
The role played by such guidelines in ensuring system safety and security
is steadily increasing in importance due to the following factors:
\begin{itemize}
\item
  the increased criticality of the software-controlled functions
  in modern systems;
\item
  the sheer complexity and number of traps and pitfalls in the most
  commonly used programming languages, such as C and C++;
\item
  the consequent ease with which programming errors are committed.
\end{itemize}
Due to this, \emph{language subsetting}, i.e.,
the prescription to only use a restricted subset of the language
such that the potential of committing possibly
dangerous mistakes is reduced,
is mandated or strongly
recommended by the most important functional safety
standards.\footnote{Such as
IEC 61508 \cite{IEC-61508:2010} (industrial, generic),
ISO~26262 \cite{ISO-26262:2018} (automotive),
CENELEC~EN~50128 \cite{CENELEC-EN-50128:2011} (railways),
RTCA DO-178C \cite{RTCA-DO-178C} (aerospace) and
FDA's \emph{General Principles of Software Validation} \cite{FDA02}
(medical devices).}
Language subsetting is generally implemented by the enforcement of
coding guidelines.

An important distinction among coding guidelines concerns those
that are only defined in terms of the actual source code and
of the toolchain used to translate it to executable code,\footnote{%
For example,
C99 has $112$ implementation defined behaviors
\cite{ISO-C-1999-consolidated-TC3}
and C18 has $119$ \cite{ISO-C-2018}.  These influence so many aspects
of source code interpretation that we can say C source code cannot be
assigned any meaning unless full details are available on
the implementation-defined behaviors of the used translation toolchain.}
and those making reference to other information, such as
requirements, specifications and designs.
The former would be amenable to full automatic checking if it was not
for \emph{undecidability}: this is a fundamental limitation
of any sufficiently-expressive programming language whereby
there is no general mechanical procedure that can decide whether or not
a program has certain properties.
\emph{Undecidable properties} are those that cannot be decided by
a general mechanical procedure, i.e., they cannot be implemented by an
algorithm.

Among the undecidable properties
are \emph{all} the more interesting ones, those that every software
engineer would want to decide: the presence or absence of buffer
or numeric overflows, invalid pointer dereferences, divisions by zero,
\dots, they are all undecidable.
As a result, guidelines that only depend on the source code and on
the language implementation are further subdivided into
\emph{decidable guidelines},
those that, at least in principle, can be verified automatically,
and \emph{undecidable guidelines}, those that cannot.

It is not difficult to give a rule of thumb for recognizing decidable
guidelines: they only depend on program properties that are
known at compile time, like
\begin{itemize}
\item
the types of the objects;
\item
the names and the scopes of identifiers;
\item
syntactic properties of the source code, like the presence of \code{goto}'s.
\end{itemize}

On the other hand, a guideline is almost certainly undecidable if it depends
on conditions that are only known at run-time, like
\begin{itemize}
\item
the values contained in modifiable objects;
\item
whether control reaches particular points.
\end{itemize}

Decidability has deep consequences on automatic analysis techniques,
particularly on static analysis.
For decidable guidelines, it is theoretically possible (i.e., modulo
the availability of sufficient computational resources)
for a tool to emit a message if and only if the rule is violated.
In contrast, for undecidable guidelines, any tool will have to implement
an approximated decision algorithm, that is, one that only in some cases
can provide a \emph{yes}/\emph{no} answer and in the remaining cases
gives a \emph{don't know} answer.
Not all tools have implementations matching this level of sophistication
(which is indeed challenging for reasons that go beyond the scope of
this paper).  In fact, several tools only provide \emph{yes}/\emph{no}
answers implying that, in reality, they are unable to recognize
the \emph{don't know} cases.  So what they do is either:
\begin{itemize}
\item
  they keep silent in cases where there can actually be a violation:
  these have false negatives and are unsuitable to safety- or
  security-related development;
\item
  they emit violation messages in cases where there may not be a violation:
  these have false positives but no false negatives, so they can be
  used for safety- and security-related development even though,
  if there are too many false positives, effectiveness of the tool
  is low;
\item
  a combination of the above, for tools having
  both false negatives and false positives.
\end{itemize}
Due also to these aspects, tool users are often confused about
undecidable guidelines and how tools report their possible violation.

In this paper, we study the relationship between undecidability
and coding guidelines in widely-used coding standards.
We focus particularly on MISRA~C, which is the most authoritative
and most widespread subset for the C programming language
\cite{MISRA-C-2012-Revision-1}.  MISRA~C, whose first edition
was published in 1998 \cite{MISRA-C-1998} and directed to the automotive
industry, has become a \emph{de facto}
standard for the development of high-integrity and high-reliability
systems in all industry sectors.
The intended audience for this paper is constituted by:
\begin{itemize}
\item
  users of coding standards and of tools supporting them;
\item
  organizations defining coding standards;
\item
  producers of tools supporting coding standards.
\end{itemize}

The paper is structured as follows:
Section~\ref{sec:undecidable-program-properties}
introduces undecidable program properties in a way that is
accessible to every software developer;
Section~\ref{sec:misra-c-and-undecidability}
illustrates undecidable MISRA~C:2012 guidelines and
classifies them according to their nature and the techniques with
which they can or cannot be checked;
Section~\ref{sec:Guidelines-on-Unreachable-and-Dead-Code}
focuses on guidelines concerning unreachable and ``dead'' code,
which have peculiarities distinguishing them from other
undecidable guidelines;
Section~\ref{sec:discussion} discusses the findings of this research
work, makes some concrete proposals for improvement, and draws
some conclusions on the desirability of undecidable guidelines
vs decidable ones;
Section~\ref{sec:conclusion} wraps up.

\section{Undecidable Program Properties}
\label{sec:undecidable-program-properties}

A function is said to be \emph{computable} if there exists an algorithm
that, given enough resources, will always produce the correct output
for each given input.\footnote{This section contains
a drastic simplification of a small part of what is presented in
any standard university course on Turing-computability.  All the mentioned
results are well known since the early 1950's and the main ideas were
established in the 1930's thanks to the work of G\"odel, Church, P\'eter,
Turing, Kleene and Post \cite{Soare16}.}
For simplicity, let us consider functions
with one input, a natural number, and a Boolean output where
we use $1$ as the representation of \emph{true} and $0$ as the
representation of \emph{false}.  The function $\fund{e}{\Nset}{\Nset}$
given by
\begin{align*}
  e(x)
    &=
      \begin{cases}
        1, &\text{if $x$ is even,} \\
        0, &\text{if $x$ is odd,}
      \end{cases} \\
\intertext{
  is clearly computable.
  It is important to observe that the condition for computability
  is the existence of the algorithm, independently from the fact
  that we have the algorithm and we know how to implement it.
  Consider the following functions:
}
  f(x)
    &=
      \begin{cases}
        1, &\text{if \emph{exactly} $x$ consecutive `5's appear}  \\
           &\text{in the decimal expansion of $\pi$,} \\
        0, &\text{otherwise;}
      \end{cases} \\
  g(x)
    &=
      \begin{cases}
        1, &\text{if \emph{at least} $x$ consecutive `5's appear}  \\
           &\text{in the decimal expansion of $\pi$,} \\
        0, &\text{otherwise.}
      \end{cases}
\end{align*}
Function $\fund{g}{\Nset}{\Nset}$ is computable: it is either
the function that always gives $1$ (if the decimal expansion of $\pi$
contains sequences of consecutive `5's of any length), i.e.,
\begin{align}
  \label{eq:g-constant-1}
  g(x)
    &=
      1, \\
\intertext{
  or there exists $k \in \Nset$ such that
}
  \label{eq:g-step-function}
  g(x)
    &=
      \begin{cases}
        1, &\text{if $x \leq k$,} \\
        0, &\text{if $x > k$.}
      \end{cases}
\end{align}
In the first case an algorithm computing $g$ is the following,
where \code{natural} is a type that encodes natural numbers:
\begin{Ccode}
  natural g(natural x) {
    return 1;
  }
\end{Ccode}
In the latter case an algorithm computing $g$ has the form
\begin{Ccode}
  natural g(natural x) {
    return (x <= k) ? 1 : 0;
  }
\end{Ccode}
The fact that we do not know yet\footnote{Unless a new result in
number theory has been published
after this paper was written.\label{fn:number-theory}}
which is the right algorithm, i.e., whether there exists $k$
such that~\eqref{eq:g-step-function} holds or, instead,
whether~\eqref{eq:g-constant-1} holds, does not really matter: $g$ is
computable.

The same thing cannot be said for function $f$:
its shape may be so complex, jumping back and forth from $0$ to $1$
in a way that no algorithm can capture,
or maybe the jump pattern is expressible by an algorithm:
we simply do not know yet.\textsuperscript{\ref{fn:number-theory}}

Let us consider a generic programming language and let $\Pset$ be the
infinite set of all its programs.  Let also $\Iset$ be the infinite set of
all inputs for the programs in $\Pset$.  For a program $P \in \Pset$ and a
possible input $I \in \Iset$, a \emph{program property} is a statement of the
form ``program $P$  [predicate] when run with input $I$.''
Examples of predicates are:
\begin{enumerate}
\item
  has 3 if-then-else's;
\item
  terminates;
\item
  divides by zero;
\item
  does not terminate.
\item
  does not divide by zero.
\end{enumerate}
For each $P \in \Pset$ and each $I \in \Iset$
let $p(P, I)$ mean ``$P$ has property $p$ when run on input $I$.''
Let us call $p_1$, \dots,~$p_5$ the properties corresponding to examples
1--5 in the enumeration above, e.g., $p_2$ is
``$P$ terminates when run on input $I$.''

For a property $p$, let us consider the decision function for $p$,
which we will denote by $\phi_p$:
it takes a program $P$ (any one in~$\Pset$), an input $I$ (any one in~$\Iset$),
and responds with $1$ if program $P$ has property $p$ when run on $I$;
it responds with $0$ otherwise.
More formally $\fund{\phi_p}{\Pset\times\Iset}{\Nset}$ is defined,
for each $P \in \Pset$ and each $I \in \Iset$, by
\begin{equation}
\label{eq:phip}
  \phi_p(P, I)
    =
      \begin{cases}
        1, &\text{if $p(P, I)$ holds,} \\
        0, &\text{otherwise.}
      \end{cases}
\end{equation}
We say that $p$ is decidable if and only if $\phi_p$ is computable.
Observe that $p_1$ is clearly decidable: an implementation of $\phi_{p_1}$
will disregard input $I$ completely, and
inspect $P$ to count the if-then-else's, returning $1$ if $P$ has
exactly $3$ of them and $0$ otherwise.

What about $p_2$?  Program termination is notoriously undecidable
for all sufficiently expressive programming languages, such as C,
Pascal, Python, and all general-purpose languages: these languages
are called \emph{Turing-equivalent} and have the property that,
if a function is computable at all, then it is computable by a program
written in any of those languages.

To see that $p_2$ is undecidable, consider the following argumentation
in a subset of the $C$ programming language where we fixed
\emph{all} the implementation-defined behaviors.
Let $\Cset$ be the subset of $C$ programs where:
\begin{enumerate}
\item
  we systematically avoid all unspecified behaviors by the use of
  temporary variables and sequence points;
\item
  we systematically define, in an arbitrary way, all undefined behaviors.
\end{enumerate}
For example, to fix the implementation-defined behaviors,
let us say something like ``we stick to the
dialect of C18 implemented by GCC version $v$ for target $z$ with
a fixed set of options
including \verb|-std=c18| and \verb|-pedantic-errors|.
For an example of point 1, we never write something like
\code{z = f() + g();}  instead, we write, e.g.,
\begin{Ccode}
  {
    T1 x = f();
    T2 y = g();
    z = x + y;
  }
\end{Ccode}
where \code{T1} and \code{T2} are the return types
of \code{f()} and \code{g()}, respectively.
For an example of point 2, we never use integer division directly, as in
\code{z = x / y;}  instead, we write, e.g.,
\begin{Ccode}
  int intdiv(int a, int b) {
    return (b == 0) ? 0 : (a / b);
  }
  ...
  z = intdiv(x, y);
\end{Ccode}
and we always use \code{intdiv()} when dividing two quantities
of (promoted) type \code{int}.
Note that we are in no way restricting the expressive power of the language:
a program relying on unspecified or undefined behavior is ill-formed anyway
\cite{ISO-C-2018}.

Now suppose, towards a contradiction, that we can actually implement
$\phi_{p_2}$, that is, we can implement in $\Cset$ a function that, given
any program $P$ and any input $I$, will give $1$ if $P$ terminates on $I$
and will give $0$ if $P$ does not terminate on $I$.
This amounts to say that we are able to complete the body of function
\code{halts()} in Figure~\ref{fig:halts} so as to implement its
specification.
\begin{figure}
  \includeCcode{halt.c}
  \caption{Source file \texttt{halt.c}, with the impossible-to-write
           function \texttt{halts()} and a main program exercising it}
  \label{fig:halts}
\end{figure}
However, if \code{halts()} can be written in $\Cset$ then it can also be
called in $\Cset$ in the way indicated in the same figure.
Then we can compile the program and execute it as follows:
\begin{verbatim}
$ gcc -std=c18 ... halt.c -o halt.exe
$ halt.exe halt.c
\end{verbatim}
For the indicated execution of \verb|halt.exe| there are only two
possibilities:
\begin{enumerate}
\item
The program prints
\begin{verbatim}
    halt.c will terminate on halt.c
    halt.c will terminate on halt.c
    ... [infinite repetitions]
\end{verbatim}
so that in fact \emph{it will not terminate!}
\item
The program prints
\begin{verbatim}
    halt.c will not terminate on halt.c
\end{verbatim}
but \emph{it has just terminated!}
\end{enumerate}
We reached a contradiction in both cases, meaning that
the function \code{halts()} cannot be written.
Note that it is not just a matter that we do not know how to write it,
we simply cannot: we will \emph{never} be able to write it.

It can be proved that \emph{all} non-trivial semantic properties of programs
are undecidable: these are all properties that depend on the contents
of writable memory locations and/or on the fact that a certain program
point can be reached or not.
Take division by zero, for instance: if we were able to implement
$\phi_{p_3}$, say, with a $\Cset$ function \code{divbyzero()}, then
we would be able to implement \code{halts()} as shown in
Figure~\ref{fig:divbyzero}.
\begin{figure}
  \includeCcode{divbyzero.c}
  \caption{Deriving a decision procedure for termination from a decision
           procedure for division by zero}
  \label{fig:divbyzero}
\end{figure}
As \code{halts()} cannot be implemented, also \code{divbyzero()} cannot.
The same holds for all non-trivial program properties: absence of
buffer overflows and any other run-time error, reachability of program points
and so on.  The proofs of these can all be obtained as variations of the
argument presented here: let the bad thing happen if and only if the
program terminates on the given input.  In this sense termination
can be called ``the father of all undecidable problems.''
However, termination is \emph{semidecidable}, meaning that the following
function \emph{is} computable:
\[
  \psi_{p_2}(P, I)
      \begin{cases}
        \mathord{} = 1,   &\text{if $P$ terminates in input $I$,} \\
        \text{undefined}, &\text{otherwise.}
      \end{cases}
\]
The algorithm is simple: run $P$ on input $I$; if and when termination
takes place, return $1$;  otherwise keep executing $P$ on $I$.
Undecidability of termination implies that, in general, we cannot
do better then that.  But consider the property $p_4$ of \emph{non-termination}:
this is not even semidecidable: in general, at no point in time are we
allowed to conclude that, having not observed termination, termination cannot
take place later.  The same holds for the property $p_5$
of \emph{non-division-by-zero}.
Worse than that, as the program input cannot typically be known in advance,
the properties we are interested in are the \emph{universal} ones, that is,
instead of $\phi_p$ of~\eqref{eq:phip}, we would need
\begin{equation}
\label{eq:upsilonp}
  \upsilon_p(P)
    =
      \begin{cases}
        1, &\text{if $p(P, I)$ holds for each $I \in \Iset$,} \\
        0, &\text{otherwise.}
      \end{cases}
\end{equation}
Of course, if $\phi_p$ is not computable, $\upsilon_p$ is also not computable.
Thus, if a property is undecidable, its universal counterpart is also
undecidable.  Take termination: its universal counterpart is called
\emph{universal termination} and a program \emph{universally terminates} if
it terminates for each input.  Universal termination not only is
undecidable, but would remain undecidable even if we had an \emph{oracle}
for ordinary termination, that is, if we had some sort of magic
behaving like $\phi_{p_2}$: this would not help as there typically are
infinitely many inputs for a program.

Finally, the fact that universal termination is undecidable, allows us
to easily prove that program equivalence is also undecidable.
Figure~\ref{fig:equiv} shows how a decision procedure for program
equivalence could be turned into a decision procedure for universal
termination.
\begin{figure}
  \includeCcode{equiv.c}
  \caption{Deriving a decision procedure for universal termination
           from a decision procedure for program equivalence}
  \label{fig:equiv}
\end{figure}

\section{MISRA Rules and Undecidability}
\label{sec:misra-c-and-undecidability}

MISRA rules are explicitly classified as \emph{decidable} or
\emph{undecidable} according to whether answering the question
\emph{``Does this program comply?''}  can be done algorithmically.
All the considerations of the previous section apply, taking into account
that all MISRA guidelines are based on \emph{universal program properties}.

Out of the 175 guidelines in MISRA~C:2012 Revision~1
with Amendment~2 \cite{MISRA-C-2012-Revision-1,MISRA-C-2012-Amendment-2},
158 are \emph{rules},
that is, guidelines such that information concerning compliance
is fully contained in the source code and in the implementation-defined
aspects of the used C language implementation.
Of the 158 rules, 37 are undecidable, of which 11 are \emph{mandatory},
22 are \emph{required}, and 4 are \emph{advisory}.\footnote{MISRA-compliant
  code must follow \emph{mandatory} guidelines: deviation is not permitted.
  MISRA-compliant code shall follow every
  \emph{required} guideline: a \emph{formal deviation} is needed
  where this is not the case.
  \emph{Advisory} guidelines are recommendations that should be
  followed as far as is reasonably practical.
  See MISRA Compliance:2020 \cite{MISRA-Compliance-2020} for details.}
Table~\ref{tab:synopsis} provides a synopsis of such undecidable rules.
It contains one row for each rule, whose identifier is written in boldface
if mandatory, in italics if advisory, or in normal font if required.
\begin{table*}
  \centering
  \caption{Synopsis of the undecidable rules in MISRA~C:2012}
  \label{tab:synopsis}
  \begin{tabular}{l|c|c|c|c|c|c|c|c|c|c|c}
    & \multicolumn{1}{p{1cm}|}{flow \mbox{undecid.}}
    & \multicolumn{1}{p{1cm}|}{numeric undecid.}
    & \multicolumn{1}{p{1cm}|}{pointee undecid.}
    & \multicolumn{1}{p{1cm}|}{side eff. undecid.}
    & \multicolumn{1}{p{1cm}|}{flow ins.\ approx.}
    & \multicolumn{1}{p{1cm}|}{type \mbox{approx.}}
    & \multicolumn{1}{p{1cm}|}{other approx.}
    & \multicolumn{1}{p{1cm}|}{coverage}
    & \multicolumn{1}{p{1cm}|}{not provable}
    & \multicolumn{1}{p{1cm}}{definition issues} \\
\hline
\MCId{R1.2}   &   &   &   &   &      &      &      &   &   & x \\
\hline
\MCId{R1.3}   & x & x & x & x &      &      &      &   &   &   \\
\hline
\MCId{R2.1}   & x &   &   &   &      &      &      & x &   &   \\
\hline
\MCId{R2.2}   & x &   &   &   &      &      &      &   & x & x \\
\hline
\MCId{R8.13}  & x &   &   &   &      &\q\q\q&      &   &   &   \\
\hline
\MCId{R9.1}   & x &   & x &   &      &      &  \q  &   &   &   \\
\hline
\MCId{R12.2}  & x & x &   &   &      &      &      &   &   &   \\
\hline
\MCId{R13.1}  & x &   &   & x &  \q  &      &\q\q\q&   &   &   \\
\hline
\MCId{R13.2}  & x &   & x & x &      &      &  \q  &   &   & x \\
\hline
\MCId{R13.5}  & x &   &   & x &  \q  &      &\q\q\q&   &   &   \\
\hline
\MCId{R14.1}  & x & x &   &   &      &      &  \q  &   &   &   \\
\hline
\MCId{R14.2}  & x & x & x &   &      &      &  \q  &   &   &   \\
\hline
\MCId{R14.3}  & x &   &   &   &      &      &      & x &   &   \\
\hline
\MCId{R17.2}  & x &   & x &   &      &  \q  &      &   &   &   \\
\hline
\MCId{R17.5}  & x &   & x &   &      &      &      &   &   &   \\
\hline
\MCId{R17.8}  & x &   & x &   &      &\q\q\q&      &   &   &   \\
\hline
\MCId{R18.1}  & x & x & x &   &      &      &      &   &   &   \\
\hline
\MCId{R18.2}  & x &   & x &   &  \q  &      &      &   &   &   \\
\hline
\MCId{R18.3}  & x &   & x &   &  \q  &      &      &   &   &   \\
\hline
\MCId{R18.6}  & x &   & x &   &      &      &  \q  &   &   &   \\
\hline
\MCId{R19.1}  & x &   & x &   &  \q  &      &      &   &   &   \\
\hline
\MCId{R21.13} & x & x &   &   &      & \q\q &      &   &   &   \\
\hline
\MCId{R21.14} & x &   & x &   &      & \q\q &      &   &   &   \\
\hline
\MCId{R21.17} & x & x & x &   &      &      &      &   &   &   \\
\hline
\MCId{R21.18} & x & x & x &   &      &      &      &   &   &   \\
\hline
\MCId{R21.19} & x &   & x &   &      & \q\q &      &   &   &   \\
\hline
\MCId{R21.20} & x &   & x &   &      &      &      &   &   &   \\
\hline
\MCId{R22.1}  & x &   & x &   &      &      &  \q  &   &   &   \\
\hline
\MCId{R22.2}  & x &   & x &   &      &      &  \q  &   &   &   \\
\hline
\MCId{R22.3}  & x &   &   &   &      &      &      &   &   & x \\
\hline
\MCId{R22.4}  & x &   & x &   &      & \q\q &      &   &   &   \\
\hline
\MCId{R22.5}  & x &   &   &   & \q\q &      &      &   &   &   \\
\hline
\MCId{R22.6}  & x &   &   &   &      &      &  \q  &   &   &   \\
\hline
\MCId{R22.7}  & x &   &   &   &      & \q\q &      &   &   &   \\
\hline
\MCId{R22.8}  & x &   &   &   &      &      & \q\q &   &   &   \\
\hline
\MCId{R22.9}  & x &   &   &   &      &      & \q\q &   &   &   \\
\hline
\MCId{R22.10} & x &   &   &   &      &      & \q\q &   &   &   \\
\hline
  \end{tabular}
\end{table*}
For each rule, information is summarized in the following columns:
\begin{description}
\item[flow undecid.]
  for rules whose undecidability directly depends on the tracking
  of control-flow;
\item[numeric undecid.]
  for rules whose undecidability directly depends on the tracking
  of numeric values;
\item[pointee undecid.]
  for rules whose undecidability directly depends on the tracking
  of pointee addresses;
\item[side effects undecid.]
  for rules whose undecidability directly depends on the tracking
  of side effects;
\item[flow ins.\ approx.]
  for rules admitting useful \emph{flow insensitive}, \emph{sound}
  approximations;
\item[type-based approx.]
  for rules admitting \emph{type-based},
  \emph{sound} and \emph{decidable} approximations;
\item[other approx.]
  for rules admitting \emph{decidable} approximations using other
  techniques;
\item[coverage]
  for rules whose compliance can only be checked by dynamic analysis
  and tracking of test coverage;
\item[not provable]
  for rules such that compliance is generally not provable,
  either programmatically or manually;
\item[definition issues]
  for rules whose definition has issues discussed in this paper.
\end{description}
In the columns labelled as approximations,
the approximability is classified using notations `\q', `\q\q' and `\q\q\q'.
The number of `\q's in the notation indicates how easy it would be
for the developers to satisfy the extra requirements due to the
approximation.
That is:
\begin{itemize}
  \item
    A single `\q' indicates an approximation having occasional violations
    whose avoidance is troublesome and/or seriously limiting the
    developers. Deviation might be appropriate.
  \item
    A `\q\q' indicates an approximation such that avoiding violations would
    need some care and possibly some rewriting, but that leads to better
    and provably correct code.  Deviation might be considered as an
    alternative to refactoring.
  \item
    A `\q\q\q' indicates an approximation where deviating violations
    is never recommendable and fixing the code is straightforward.
\end{itemize}

Consider the following MISRA~C:2012 rule:
\begin{quote}
\MCId{R22.5}: \MCHd{R22.5}
\end{quote}
The reason why this rule is \emph{flow undecidable} is its \emph{flow
sensivity}, meaning that the rule is not violated if a pointer to a
\code{FILE} object is dereferenced in unreachable code:
\begin{Ccode}
  FILE *p;
  /* ... */
  if (always_false_in_this_configuration(/* ... */) {
    FILE f = *p;  // Unreachable: not a violation.
  }
\end{Ccode}
The obvious \emph{flow-insensitive}, decidable approximation consists
in flagging all dereferences of pointers to \code{FILE} objects independently
from reachability.
In this and other cases, \emph{flow sensitivity}
(ubiquitous in undecidable rules as you can see in Table~\ref{tab:synopsis})
does more harm than good: in the example
above, the reason that \MCId{R22.5} is not violated is because the \emph{then}
branch is unreachable. What is the point of this exemption?
In fact the unreachable code is in violation of \MCId{R2.1}
(discussed in Section~\ref{sec:Guidelines-on-Unreachable-and-Dead-Code}).

For another example, consider the following guideline:
\begin{nbquote}
\MCId{R17.8}: \MCHd{R17.8}
\end{nbquote}
This is undecidable for two reasons.
The first is, again, \emph{flow sensitivity}: a modification happening
on a branch that is not reached is not a violation.
Consider the following
\begin{example}
\begin{Ccode}
  void f(uint32_t x) {
    if (x < 0) { // If always x >= 0 on entry...
      x = 0;     // ... Rule 17.8 is not violated.
    }
    /* ... */
\end{Ccode}
\end{example}
\pagebreak
Flow sensitivity is not the only cause of undecidability
for \MCId{R17.8}.  Consider the following
\begin{example}
\label{ex:MC3R1.R17.8-flow-insensitive}
\begin{Ccode}
  extern void g(uint32_t *p);

  void f(uint32_t x) {
    /* ... */
    g(&x); // Rule 17.8 violation?
    /* ... */
\end{Ccode}
\end{example}
Knowing whether \MCId{R17.8} is violated depends on knowing whether
function \code{g()} modifies the pointee of the argument
it received on input, which also depends on the tracking of pointee
addresses, and this is undecidable.
In this and other cases, the authors of this paper believe that the latitude
allowed by undecidability has insufficient rationale.
This is why, for several rules, Table~\ref{tab:synopsis} goes beyond
crossing \emph{flow insensitive approximation} by crossing
\emph{type approximation}.  We call ``type approximation'' one
that can be expressed in a stricter type system than the standard
C type system.\footnote{This notion is not new to MISRA C practitioners:
  MISRA~C:2012 \emph{essential type model} along with the guidelines
  that are based on it, define a type system which is stronger than
  the C type system.}
Note that static type approximations are, by definition, flow insensitive
and decidable.\footnote{\emph{Static typing} is in contrast with
\emph{flow-sensitive typing}, where the type of expressions may
depends on their position in the control flow.
In flow-sensitive type systems, the type of an expression may be updated
to a more specific type following an operation validating the subtype.
For example, just after \code{p = malloc(sizeof(int))} the type of
\code{p} may be an encoding of ``null pointer or pointer to the beginning
of a block in the heap'', but within the \emph{then} branch of a subsequent
if-then-else guarded by \code{p != NULL}, the type of \code{p}
may be updated to ``pointer to the beginning of a block in the heap.''
The Rust language, which we will mention later in the paper,
is based on flow-sensitive typing \cite{KlabnikN18}.}
Concerning \MCId{R17.8}, a sensible type approximation would go
along the lines ``Function parameters should be considered as read-only.''
A static analyzer can clearly check this, in particular by checking
that no explicit and implicit casts can circumvent the writing prohibition,
and flag the violation in Example~\ref{ex:MC3R1.R17.8-flow-insensitive}.

We will now go through other rules in Table~\ref{tab:synopsis},
explaining their classification and their relationship
with undecidability.

\begin{quote}
\MCId{R1.2}: \MCHd{R1.2}.
\end{quote}
The rationale of \MCId{R1.2} is that programs relying on extensions
will be difficult to port to a different language implementation.
In addition, extensions make \emph{compiler qualification},
as mandated by functional safety standards, more expensive,
as it requires in-house development of test cases for
the extensions~\cite{BagnaraBH22}.

An interesting thing about \MCId{R1.2} is that it is the only
MISRA~C:2012 rule that is tagged both as \emph{Undecidable}
and \emph{Single Translation Unit}, meaning that all violations
involve a single translation unit only, i.e., the compiler
and not the linker.  However, the point with language extensions
is that, by definition, they are not known in advance.
Consider the following C fragment:

\pagebreak
\includeCcode{exten.c}

This contains an empty initializer and returning of a \code{void} expression,
which are undefined in all versions of the C standard.
Nonetheless the GCC documentation does not document
them as extensions.\footnote{See
\url{https://gcc.gnu.org/onlinedocs/gcc/C-Extensions.html},
last accessed and checked on October 6, 2022.}
Can an undocumented compiler feature be accepted as a legitimate
language extension?  Probably not: as the presence of documentation
is crucial and checking that is a human activity, \MCId{R1.2} should
probably be a directive.

\begin{quote}
\MCId{R1.3}: \MCHd{R1.3}
\end{quote}
\MCId{R1.3} covers all undefined and critical unspecified
behaviors that are not covered by other rules: there are many
of them, so the crosses given in the corresponding row in
Table~\ref{tab:synopsis} represents a summary.

\begin{quote}
\MCId{R8.13}: \MCHd{R8.13}
\end{quote}
The reason why \MCId{R8.13} is undecidable is that the missing
\texttt{const} qualification might impact on code that is unreachable.
As unreachable code is flagged by \MCId{R2.1} (see
Section~\ref{sec:Guidelines-on-Unreachable-and-Dead-Code}),
the rationale for accepting undecidability is weak.
We believe the stronger, decidable version, would be more useful
without constraining programmers too much; hence in Table~\ref{tab:synopsis},
in the column labeled \emph{type-based approx.}, the rule has `\q\q\q'.

\begin{quote}
\MCId{R9.1}: \MCHd{R9.1}
\end{quote}
Missing initialization of automatic variables is the origin
of many defects and vulnerabilities.
\MCId{R9.1} is undecidable because it rules out
reading uninitialized stack cells: capturing this with high
precision in static analysis is challenging, especially
when arrays are involved.  The rule, however, has
sensible decidable approximations.
The simplest one is to ``always initialize automatic variables
at declaration time.''
This is less extreme than it might seem:
MISRA~C:2012 itself hints at this direction when defining
\MCId{R2.2} (which will be examined in
Section~\ref{sec:Guidelines-on-Unreachable-and-Dead-Code}),
by specifiying that initializations may be kept even when redundant.
Most importantly, wholesale initialization of automatic variables
is now optionally implemented by major compilers (GCC from version 12, Clang
from version 8), with at most negligible slowdowns and speedups
in some cases \cite{Bastien19,CookZ21,Zhao21}.\footnote{%
The reason for the occasional speedups sems to be that systematic
zero-initialization improves superscalar execution in the CPU due
to the breaking of dependencies.}
It is interesting to note how, in front of an important rule that is targeted
by basically all bug finders (with false negatives and/or false positives
due to the rule undecidability), a very speed-sensitive community like
the one revolving around the Linux kernel is seriously considering
the systematic use of such options.

\begin{quote}
\MCId{R14.1}: \MCHd{R14.1}
\end{quote}
This rule, together with \MCId{R14.2}, aims at introducing in C
a sort of \emph{determinate iteration} construct like Pascal's
\code{FOR} looping construct.\footnote{A \emph{determinate iteration}
  construct is one such that the maximum number of iterations
  is known before the first iteration begins.  None of the looping constructs
  of C have this property.}
This ideal, which is not completely achieved, rests on restrictions
that are applied to C's \code{for} loops.
As the restrictions are semantic in nature, both \MCId{R14.1} and \MCId{R14.2}
are undecidable.  The type restriction given in the headline of \MCId{R14.1}
might be surprising, as static type restrictions are decidable.
The undecidability is due to the specification of \emph{loop counter}: a
loop counter is variable that satisfies three conditions, two of which are
undecidable \cite[Section~8.14]{MISRA-C-2012-Revision-1}.
Syntactic, fully decidable restrictions are of course possible and
would have the advantage of fully achieving the goal of having
determinate iteration in C.

\begin{quote}
\MCId{R17.2}: \MCHd{R17.2}
\end{quote}
There are two reasons why this rule is undecidable: flow sensitivity
and function pointers.  A variant that is flow insensitive and
forbids the use of function pointers would be decidable.  When function
pointers are needed to implement, e.g., callbacks, a type-based approximation
(effectively limiting the use of function pointers so that recursive
calls via them are impossible) would be decidable and would
cover most cases that occur in embedded system programming.

\begin{quote}
\MCId{R21.13}: \MCHd{R21.13}
\end{quote}
Undecidability of this rule comes, besides flow-sensitivity, from the
undecidability of value tracking.  A type-based approximation,
where the static analyzer enforces constraints on a type only
capable of representing the value \code{EOF} and those representable
by an \code{unsigned char}, provides a decidable way of ensuring compliance.

\begin{quote}
\MCId{R21.14}: \MCHd{R21.14}
\end{quote}
This is undecidable for the same reasons as \MCId{R21.13} and the same
discussion applies.  Indeed, null-terminated strings deserve not just
a fictitious type that only manifests itself within the static analyzer:
they deserve an explicit \code{typedef} name so that developers are very
aware when they manipulate null-terminated strings and not ordinary
character arrays.

\begin{quote}
\MCId{R21.19}: \MCHd{R21.19}
\end{quote}
The very same approach described for \MCId{R17.8} can be used.\footnote{%
This rule might be extended to also cover
\code{strchr()}, \code{memchr()} and similar functions
whose indiscriminate use can circumvent the \code{const} promise
of their parameter.}

\begin{quote}
\MCId{R22.1}: \MCHd{R22.1}
\end{quote}
For this rule as well as \MCId{R22.2} it is possible to use
an \emph{ownership} model similar the one implemented in
Rust \cite{KlabnikN18} and based on \emph{flow-sensitive typing}.

\begin{quote}
\MCId{R22.3}: \MCHd{R22.3}
\end{quote}
The problem with this rule is that it cannot be checked automatically
well beyond the problem of undecidability.  The notion
of being ``the same file'' is not definable at the C source level
even taking into account the implementation-defined behaviors of
the implementation.
Such notion depends on peculiarities of file systems and on their
current state: relative paths, hard links, symbolic links, logical drives
and other file system features are such that the check for compliance
requires a lot of information that only knowledgeable humans can provide.
In other words, this MISRA guideline should probably be a directive.

\begin{quote}
\MCId{R22.7}: \MCHd{R22.7}
\end{quote}
This rule is undecidable because of flow sensitivity
and because of the semantic notion of \emph{unmodified value}.
A type-based approximation drastically restricting what can be
done on the return values of the indicated functions would be decidable.

\begin{quote}
\MCId{R22.8}: \MCHd{R22.8}
\end{quote}
This rule, as for the associated \MCId{R22.9} and \MCId{R22.10},
is undecidable because it is overly permissive in the forms and
positioning of the operations that zero and test \code{errno}.
Syntactic restrictions would result into decidable guidelines
without constraining the programmer in an unacceptable way.

\section{Guidelines on Unreachable and Dead Code}
\label{sec:Guidelines-on-Unreachable-and-Dead-Code}

MISRA~C:2012 has two required rules to deal with useless, and thereby
possibly undesirable, code.  The first one is:

\begin{quote}
\MCId{R2.1}: \MCHd{R2.1}
\end{quote}

There, \emph{unreachable code} is code that cannot be executed
unless the program has undefined behavior.
The rationale of the rule is that the presence of unreachable code
may:
\begin{itemize}
\item Indicate an error in the program's logic: as the compiler
  is allowed to remove unreachable code (but it is not
  required to do so),\footnote{Even though not explicitly mentioned,
    the linker (and in some cases \emph{only} the linker), can also
    remove unreachable code from the executable, while not being
    required to do so.} why is the code there?
\item Waste resources and prevent optimizations in the case the compiler
  and the linker do not remove the unreachable code.
\end{itemize}
\pagebreak
To this, we can add that unreachable code that ends up in the executable
program is, per se, a security issue: an attacker might exploit another
vulnerability in order to actually reach that code and achieve
its malicious goals.

Compliance with \MCId{R2.1} cannot be proved by means of static analysis
alone.\footnote{Experts might object that \emph{symbolic model checking} may
be sufficient in some cases.  However, since this is based on the synthesis
of test cases that are validated by \emph{symbolic} or \emph{concolic}
(\emph{conc}rete mixed with symb\emph{olic}) execution,
we assimilate this technique to dynamic analysis.}
Static analysis can reveal that some functions are never called because,
e.g., they never occur in explicit function calls and their addresses are
never taken;  the user can supplement this information by annotating
functions that are only called via assembly code or as interrupt services
routines.  Static analysis can also reveal that a certain condition is
always true or always false, when this does not depend on external inputs,
or that an expression used to control a \code{switch} statement can
only take certain values.  But, in general, making sure all code is
actually reachable requires dynamic analysis: 100\% statement coverage
has to be reached for that purpose, with a test-suite that is
\emph{a subset of the input space for the program}.
Note that we are not talking of \emph{unit testing} here (reaching 100\%
statement coverage via unit tests does not ensure compliance), but rather
of \emph{system testing}.

\begin{quote}
\MCId{R2.2}: \MCHd{R2.2}
\end{quote}

MISRA~C:2012 \MCId{R2.2} is a required rule that defines \emph{dead code} as
``Any operation that is executed but whose removal would not affect
program behaviour,'' further specifying that
``unreachable code is not dead code as it cannot be executed.''\footnote{%
The notion of \emph{operation} is not explicitly defined in MISRA~C:2012
or in the C Standard.  We interpret \emph{operation} as to signify
``anything that
in the C Standard is called \emph{operation}: integer operations,
floating-point operations, arithmetic operations, bitwise operations,
atomic operations, synchronization operations,
pointer operations, \dots''.}
Three citations from functional safety standards are provided:
IEC~61508-7 Section~C.5.10 \cite{IEC-61508-07:2010},
DO-178C Section~6.4.4.3.c \cite{RTCA-DO-178C}, and
ISO~26262-6 Section~9.4.5 \cite{ISO-26262-06-2018}.
The first two references are also cited for \MCId{R2.1}, and in fact:

\begin{description}
\item[IEC 61508-7 Section C.5.10]
This section does not define \emph{dead code}, but gives a terse introduction
to data-flow analysis.  It gives three examples of use, two of which
do overlap with the MISRA~C:2012 definition of \emph{dead code}: these
concern values being unnecessarily written to memory, a.k.a.
\emph{dead stores} in other literature.
\item[DO-178C Section 6.4.4.3.c]
Section 6.4.4.3 is titled ``Structural Coverage Analysis Resolution'',
which already hints at the fact that MISRA C:2012 definition of dead
code is not compatible with DO-178C definition.
Such definition can be found in Annex B, Glossary:
``Dead code --- Executable Object Code (or data) which exists as a result
of a software development error but cannot be executed (code) or used
(data) in any operational configuration of the target computer
environment. It is not traceable to a system or software requirement.''
DO-178C also defines a different notion of ``deactivated code.''
\item[ISO 26262-6 Section 9.4.5]
This section of ISO 26262-6 is concerned with structural coverage.
To exemplify the use of structural coverage methods, two examples
are given:
\begin{quote}
``EXAMPLE 1
Analysis of structural coverage can reveal shortcomings in
requirements-based test cases, inadequacies in requirements, dead
code, deactivated code or unintended functionality.''
\end{quote}
\begin{quote}
``EXAMPLE 2
A rationale can be given for the level of coverage achieved based on
accepted dead code (e.g. code for debugging) or code segments
depending on different software configurations; or code not covered
can be verified using complementary methods (e.g. inspections).''
\end{quote}
So, while ISO 26262 does not give a definition of \emph{dead code},
it clearly states that dead code can be found by analyzing structural
coverage.
\end{description}

Summarizing, the definition of \emph{dead code} given in MISRA~C:2012 is not
compatible with the definitions given in ISO~26262 and DO-178C.
The only overlap between the functional safety standards cited by
MISRA~C:2012 and its notion of \emph{dead code} is given by
\emph{dead stores}.  In other words, the named standards
define \emph{dead code} as \emph{code that cannot be executed
plus data that is written and cannot be read}.  This is a much
narrower definition than the MISRA~C:2012 definition
as ``operations that can be removed without affecting program behaviour.''

MISRA~C++:2008 has the following required
\begin{quote}
  Rule 0-1-9 There shall be no dead code.
\end{quote}
This is different from MISRA~C:2012 \MCId{R2.2} in that \emph{dead code}
is characterized as ``any executed statement whose removal would not affect
program output [...]''.
In the sequel, in order to avoid confusion with the definitions used
in functional safety standards, we will use the expression
\emph{effectless code} instead of the MISRA~C:2012 and the
MISRA~C++:2008 notions of \emph{dead code}.

In the online version of the
\emph{SEI CERT~C Coding Standard}\footnote{%
\url{https://wiki.sei.cmu.edu/confluence/display/c/SEI+CERT+C+Coding+Standard}, last accessed on October 4, 2022. Note that
\textit{Recommendation MSC12-C} is not contained in the printed version
\cite{CERT-C-2016}.}
this topic is covered by the following \emph{Recommendation}:
\begin{quote}
  \textit{MSC12-C} Detect and remove code that has no effect or is never executed
\end{quote}
SEI CERT~C Coding Standard uses the word \emph{Recommendation} as opposed
to \emph{Rule}: whereas rules are normative (though not necessarily
amenable to automatic analysis), ``recommendations are meant to
provide guidance that, when followed, should improve the safety,
reliability, and security of software systems.''
Following MISRA terminology, CERT-C \textit{Recommendation MSC12-C}
would be an advisory directive, whereas MISRA~C:2012 \MCId{R2.2}
is a required rule.

A wholesale ban on effectless code, like the one required by
MISRA~C:2012 \MCId{R2.2}, discourages sane things like the ones shown
in Figure~\ref{fig:undesirable-MC3R1.R2.2-violations-examples}.
\begin{figure*}
\begin{Ccode}
x + OFFSET;     // Addition is justifiable, even when OFFSET is defined to be 0.
x * SCALE;      // Multiplication is justifiable, even when SCALE is defined to be 1.
x * sizeof(T);  // Multiplication is justifiable, no matter what the value of
                // sizeof(T) is.
do_X_if_necessary(); // The function may be inline and its body may be empty
                     // (e.g., #ifdef-ed out) in this configuration.

typedef enum {
  BIT0 = 1U << 0,  // Justifiable (no-op) shift by 0 positions.
  BIT1 = 1U << 1,
  BIT2 = 1U << 2
} Bit_Masks;
\end{Ccode}
\caption{Examples of undesirable violations of MISRA~C:2012 Rule~2.2}
\label{fig:undesirable-MC3R1.R2.2-violations-examples}
\end{figure*}

Even a wholesale ban on dead stores is undesirable.
Consider the following example
\begin{Ccode}
void transmit_octet(const uint8_t octet) {
  uint8_t mask = 1U;
  for (uint8_t bit = 0; bit < 8; ++bit) {
    transmit_bit(octet & mask);
    mask <<= 1U;
  }
}
\end{Ccode}
The shift-assignment to \code{mask} on the last iteration is a dead store.
However, modifying the source code in order to avoid it would almost
certainly decrease code quality for no gain, not even on efficiency.
The point is that the rationale against dead stores is quite weak:
while it is true that they may indicate a programming mistake, often
they do not, and compilers are quite good at detecting them and optimizing
them out when there is incentive to do so.

An important issue with MISRA~C:2012 \MCId{R2.2} is that it is the
only undecidable MISRA~C:2012 rule that is also unprovable.
There is neither a static or dynamic analysis, nor a sensible review
process that can decide whether an operation can be removed from a
program without affecting its behavior.

Note the difference with respect to Rule~2.1: achieving 100\% statement
coverage proves that a project is compliant.  Moreover, for MISRA~C:2012

\begin{quote}
\MCId{R14.3}: \MCHd{R14.3}
\end{quote}
achieving 100\% branch coverage proves compliance.
Of course, static analysis techniques can pinpoint some instances
of non-compliance with these rules: what remains to be proved
for compliance can be done via dynamic analysis.
In contrast, for \MCId{R2.2}, proving compliance is generally impossible:
it is not just that program behavior equivalence is strongly undecidable,
as we saw in Section~\ref{sec:undecidable-program-properties}.
In order to prove compliance with \MCId{R2.2}, one should, for each
combination of operations in the program (the number of which is
exponential in the total number of operations), prove that
removing that combination preserves program behavior.
In other words, proving compliance with respect to \MCId{R2.2}
would require answering with \emph{no} an exponentially large
number of questions (exponential in the size of the program)
of the form ``is the transformed program, where we have deleted some operations,
behaviorally equivalent to the original program?''
Answering these question programmatically cannot be done,
it cannot be done via dynamic analysis, and it cannot be done
manually as program equivalence requires the same behavior
for each of the possible inputs, of which there typically is
an infinite number.

Guidelines for which compliance is practically impossible to be proved
serve no purpose.  For instance, when confronted with rule
MISRA C:2012 \MCId{R2.2}, users will either:
\begin{enumerate}
\item
  unknowingly fake compliance (possibly with the complicity of tool vendors
  making claims that they cannot actually make about the coverage of the rule);
  or,
\item
  when they are knowledgeable enough, they will raise a project
  deviation saying that they did their best and that they have confidence
  that remaining effectless code, if any, is not causing problems.
\end{enumerate}

\section{Discussion}
\label{sec:discussion}

In this section we briefly discuss the tradeoff
between decidable and undecidable guidelines and we put forward
a concrete proposal about the treatment of effectless code.

\subsection{How Good Are Undecidable Coding Guidelines?}

Are undecidable coding guidelines good or bad?
The attentive reader already knows the position of the authors:
\begin{itemize}
\item
  an undecidable guideline is good when there are no decidable
  sound approximations for it (that is, one that catches
  all its violations plus more) or when such decidable approximations
  would tie the hands of programmers in a way they cannot
  easily achieve their objectives;
\item
  in all other cases, undecidable guidelines are bad.
  \end{itemize}
Undecidable guidelines are troublesome because they let programmers
deal with false positives and/or false negatives: if there are no
false negatives then there may be many false positives and these
are time consuming to deal with;  if there are no or few false positives,
then there may be false negatives, in which case developers will
have to look for different or additional solutions.  Note that this
is a universal constant: if a guideline is undecidable, any fully
automatic checker will have false positives or false negatives or both.

It is very instructive to observe that, independently from the
number of companies and organizations that offer verification
tools basically for free to the Linux community:
\begin{enumerate}
\item
  for the undecidable undefined behavior caused by reading
  uninitialized automatic variables, as we have already observed, they
  are looking at compiler options to solve the problem;
\item
  for the various undecidable undefined behaviors related to memory
  management errors, they are looking at Rust
  \cite{Salter21,Cantrill19,Wallen21,Vaughan-Nichols21a,Vaughan-Nichols21b,Tung21,Melanson21,Elhage20}.
\end{enumerate}
While point 1 involves no effort on the part of the developers,
point 2 and the possible move to Rust require a lot of discipline
on the part of the programmers, which shows that the Linux community,
when it comes to important safety and security matters, is in line
with MISRA~Compliance:2020 where is says that simply satisfying
the immediate convenience of the developer does not constitute
an acceptable rationale for deviating guidelines
\cite[Section~4.4]{MISRA-Compliance-2020}.

Note that we are not necessarily proposing to change the rules
in Table~\ref{tab:synopsis} so as to make all those with `\q\q' or
`\q\q\q' in at least one \emph{approx.} column decidable.
Another possibility is for tools to implement some sound decidable
approximations for those rules

\subsection{An Alternative To the Strict Ban on Effectless Code}

As we have seen in
Section~\ref{sec:Guidelines-on-Unreachable-and-Dead-Code}, in general
there is no way to actually, fully comply with MISRA~C:2012 \MCId{R2.2}
or CERT-C \textit{Recommendation MSC12-C}.
For projects with MISRA-compliance requirements this is a problem:
as they cannot claim compliance with MISRA~C:2012 \MCId{R2.2}
or MISRA~C++:2008 Rule 0-1-9, they will have to deviate.
For the relatively few cases that a tool can detect:
\begin{itemize}
\item
  the code can be amended if this has a positive effect on code quality;
\item
  a deviation has to be raised otherwise, as recommended by
  MISRA~Compliance:2020 \cite{MISRA-Compliance-2020}
  (code quality always comes first).
\end{itemize}
In any case, a deviation will have to be raised \emph{in addition}
to all that, simply because nobody can know whether there are other
undetected violations of the rule.  Unfortunately, this does not match any of the
allowed rationales for deviation allowed by MISRA~Compliance:2020
(code quality, access to hardware, integration or
use of suitably qualified adopted code).
Pragmatically, projects will have no choice other than raising
a project deviation with a justification along with the following
lines:
\begin{quote}
  Peer review gives us confidence that no evidence of errors in the
  program's logic has been missed due to undetected violations of
  \MCId{R2.2}, if any.  Testing on time behavior gives us confidence
  on the fact that, should the program contain dead code that is not
  removed by the compiler, the resulting slowdown is negligible.
\end{quote}

A possible solution to rectify the situation is by replacing
MISRA~C:2012 \MCId{R2.2} with a directive whose headline can
be something like
\begin{quote}
  Unjustified effectless code shall be minimized.
\end{quote}
Note the similarity with required directive
\begin{quote}
\MCId{D4.1}: \MCHd{D4.1}
\end{quote}
\MCId{D4.1} takes a very pragmatic approach to a much more serious problem
than effectless code: this is reasonable as ensuring the absence of
run-time errors is impossible as is ensuring the absence of effectless code.

The notion of \emph{unjustified} can be defined as follows:
code is \emph{unjustified} if it has both of the following attributes:
\begin{enumerate}
\item
it does not help understanding of the algorithm;
\item
it does not come from the natural abstraction of the algorithm
so that it can be applied in different situations (e.g., on different
architectures and/or on different configurations).
\end{enumerate}
In particular effectless code is justified if it arises from an abstraction
process.  That can be:
\begin{itemize}
\item
  data abstraction: macro expansions, macro definitions;
\item
  control abstraction: loops, recursion.
\end{itemize}
\begin{figure*}
\begin{Ccode}
  x + 0;          // Unjustified addition.
  x + OFFSET;     // Justified addition, even when OFFSET is defined to be 0.

  x * 1;          // Unjustified multiplication.
  x * SCALE;      // Justified multiplication, even when SCALE is defined to be 1.
  x * sizeof(T);  // Justified multiplication, no matter what the value of sizeof(T) is.

  do_X_if_necessary();  // Justified function call, unless it can be argued that for all present and future
                        // project configurations, the function has no influence on the program behavior.

  for (i = 0; i < NUM_REPETITIONS; ++i) { // Entire loop justified, even if NUM_REPETITIONS
    do_things();                          // expands to 0 or 1 in this configuration.
  }


  // Saturate.
  x = (x > MAX) ? MAX : x;  // Justified, even if X is never greater than MAX in this configuration.
\end{Ccode}
\caption{Examples of justified and unjustified effectless code}
\label{fig:proposed-directive-examples}
\end{figure*}
Figure~\ref{fig:proposed-directive-examples} provides examples of
compliance and non-compliance to this hypothetical directive.

Compliance to the directive requires planning and documenting
activities, possibly involving static analysis and dynamic analysis
techniques, along the lines of \MCId{D4.1} (i.e., with reference to design
standards, test plans, static analysis configuration files and code
review checklists).

\section{Conclusion}
\label{sec:conclusion}

Undecidability is an inescapable limitation of computing whereby
only purely-syntactic properties of programs written in languages
such as C and C++ are algorithmically verifiable or refutable:
all other properties are undecidable,
meaning that there will never be a general algorithm that can decide
whether a program has or does not have the property.
As a result, most of the program properties associated to program safety
and security requirements are undecidable.

Coding guidelines that embody the language-subsetting requirement
of many functional safety standards are constrained between two
conflicting goals:
\begin{enumerate}
\item
  they have to prevent bad things from happening;
\item
  they have to be acceptable to developers.
\end{enumerate}
Goal~2 implies the coding guideline should be directly targeted at
preventing the bad thing.  Given that the possibility of
occurrence of the bad thing is usually undecidable, the conjunction
of goal~1 with goal~2 tends to favor undecidable guidelines.
The tradeoff changes between communities and with time:
\begin{itemize}
\item
  developers of critical software in highly-regulated industry sectors
  are more willing to exchange a little bit of inconvenience with
  the strong guarantees that decidable guidelines can provide;
\item
  even highly-unregulated communities, like the one revolving around
  the Linux kernel, seem now inclined to accept more restrictions:
  something that, only a few years ago, would have been vehemently
  rejected.
\end{itemize}

While undecidable coding guidelines cannot pragmatically
dispensed with, at least for languages like C and C++,
analysis tools can, in several cases, be based on some decidable
approximations.
Nonetheless, undecidable program properties tend to confuse developers,
no matter whether the guidelines used to ensure or exclude
the program has the property are undecidable or decidable.
Many developers are distracted by (sometime false) thoughts
like ``There is no problem in my program, why this violation?''
without considering that no perfect solution exists and that,
by necessity, we need to compromise.

We believe this paper, in which we studied the role played
by undecidability in the most widely used coding standard,
MISRA~C, will help developers in better appreciating what the problems
are and which tradeoffs have to be faced.
Some of the findings of this research were totally unexpected
at the outset: as a result, this work goes beyond its original
survey/educational goals by uncovering some real problems and
corresponding possible solutions.

\subsection*{Acknowledgments}

The MISRA~C Guideline headlines are reproduced with permission of
\emph{The MISRA Consortium Limited}.
We are grateful to Federico Serafini (University of Parma and BUGSENG)
for his comments on the draft versions of this paper.

\providecommand{\noopsort}[1]{}

\end{document}